# Practical random number generation protocol for entanglement-based quantum key distribution

G. B. Xavier[*], T. Ferreira da Silva, G. Vilela de Faria, G. P. Temporão

and J. P. von der Weid

*Center for Telecommunication Studies, Pontifical Catholic University of Rio de Janeiro, R. Marquês de São Vicente 225 – Rio de Janeiro - Brazil*

*Corresponding author: guix@opto.cetuc.puc-rio.br

A simple protocol which takes advantage of the inherent random times of detections in single photon counting modules is presented for random active basis choices when using entanglement-based protocols for Quantum Key Distribution (QKD). It may also be applicable to the BB84 protocol in certain cases. The scheme presented uses the single photon detectors already present on a QKD setup, working on the same rate as the system is capable of detecting, and is, therefore, not limited by the output rates of quantum random number generators. This protocol only requires small hardware modifications making it an attractive solution. We perform a proof-of-principle experiment employing a spontaneous parametric down-conversion process in a $\chi^{(2)}$ non-linear crystal to demonstrate the feasibility of our scheme, and show that the generated sequence passes randomness tests.

## 1. Introduction

Quantum Key Distribution (QKD) [1] is an elegant solution to distribute a secret key between two remote parties with absolute security. It is also one of the first practical applications at the levels of the single quanta [2]. Experimental progress was shy in the early days of the technology, but it is quickly becoming more and more practical, with major improvements being developed [3]. One of the key features for QKD security is the need of Random Number Generators (RNGs) that the parties use to choose the basis in which they measure the transmitted bit [1]. In this paper we present a simple protocol to be used along with an active basis choice BBM92 [4] QKD scheme which can work at any key detection rate, since it is only dependent on the detection hardware, and thus not limited by the speeds of existing Quantum Random Number

Generators (QRNGs) [5-8]. It also dismisses their usage, thus reducing the costs. Our proposal relies on the randomness of the time between detections of the Single Photon Counting Modules (SPCMs), already present in any QKD system, requiring small modifications to the setup. Our idea can be used in the E91 protocol [9], which also employs entangled pairs of photons. Finally it may also be used by Bob (the receiver), and in some cases also by Alice (the transmitter) in the case of the BB84 [1] protocol.

All QKD protocols rely on the ability to generate true random numbers. In the BB84 protocol, for example, Alice randomly creates a classical bit string to be transmitted, and for each bit she also randomly selects between two distinct maximally overlapped bases for qubit encoding. Bob, randomly and independently from Alice, also chooses for each incoming qubit the same two bases. In order to perform truly random and independent choices, a hardware-based true QRNG must be used by both Alice and Bob, as any software RNG actually generates a pseudo-random sequence [10-11].

All of the existing hardware-based solutions can provide truly random numbers, although they are limited in the bit generation rate. As QKD systems evolve into faster rates [12] deployed QRNGs will need to keep up. Our proposal is independent of the QRNG bit generation rate because it uses the same detectors already in use in a QKD setup.

Whereas BB84 requires active participation of Alice on the choice of the binary digits that will comprise the secret key, some other protocols, like BBM92, which employs pairs of entangled photons [13-15], require random numbers only for basis choices. In this case, an entangled photon-pair source is placed somewhere between Alice and Bob, and one photon from the pair is sent to each one. Similarly to the BB84 case, they measure their photons using two randomly selected bases. The other main entanglement based QKD protocol is E91, and the main difference to BBM92 is that three basis are needed by both Alice and Bob, with this additional basis required for verification of the CHSH (Clauser-Horne-Shimony-Holt) inequality to ensure unconditional security [16]. We shall keep referring only to the BBM92 case for entanglement-based protocols until the end of the paper as it is simpler to implement. In both protocols unconditional security can only be obtained if the basis choices performed by Alice and Bob are done independently and randomly, otherwise an eavesdropper (Eve) can, in principle, predict some of the bases used and gain partial information about the key.

Recent experiments on the BBM92 protocol used a passive basis choice for Alice and Bob (a 50/50 % beam splitter for example) [17,18], which is indeed a truly random choice. However, four detectors at Alice and at Bob are needed to fully implement the protocol, making it a costly solution, especially if one wishes to work in the 1550 nm telecom window, where expensive detectors are normally used, such as InGaAs Avalanche Photon Detectors (APDs) [1] or, more recently, superconducting nanowire detectors [3,19]. An active choice scheme may be performed if an element is used to intentionally change the measurement basis, e.g., a half-wave plate or a phase modulator in the cases of polarization or phase coding respectively, reducing the number of detectors to two at both Alice and Bob. In this case the user must provide the random choice in order for the protocol to remain secure, which is typically done through an external hardware-based QRNG. We now present a scheme that dismisses the use of additional QRNGs for active basis selection in the BBM92 protocol, which may also be extended to Bob and in some cases to Alice, for basis selection in the BB84 protocol. This scheme takes advantage of the fact that the time between detections of single photons are random, and follow an exponential probability distribution [5].

## 2. The protocol

In our proposal for the BBM92 protocol both Alice and Bob have a clock generator each, working asynchronously from each other. This signal can be generated within modern electronics already used in a QKD setup. The main idea is to create the random numbers based on the number of clock pulses between consecutive photon detections, that is, between consecutive generations of electrical pulses in the output of their detectors. Since the entangled source may not be trusted, the following procedure is performed: initially, both Alice and Bob block their detector inputs and wait for the first dark count, thus obtaining random and independent integers equal to the number $N_{A0}$ ($N_{B0}$ for Bob) of clock pulses until either detector has fired. Alice (Bob) will proceed to calculate $N_{A0}$ mod 2 ($N_{B0}$ mod 2) and choose one of the 2 needed bases for the first transmitted qubit depending on the result. They keep performing the procedure The detector inputs are opened, the quantum transmission starts and they count the number of pulses $N_{A1}$ ($N_{B1}$) until the next detection, perform $N_{A1}$ mod 2 ($N_{B1}$ mod 2), choose the basis for the next qubit and so on, where a random number $N_{AT(BT)}$ will be obtained between detections $T-1$ and $T$. Since the times of detection follow an exponential distribution [7,8], the generated sequence will be random. As long as we can assume Eve is not looking inside the detectors (however, Eve can still "hear" the detection clicks without gaining any information, in exactly the same way as in the standard BB84 protocol), she will not be able to guess the bases used, as required in all QKD protocols. The scheme is presented in Fig. 1a for the particular case of the BBM92 protocol with polarization encoding in optical fibers, which is a feasible option now that active fully continuous real-time polarization stabilization for QKD has been demonstrated [20].

This same idea can also be extended to the BB84 protocol from the point of view of Bob, since he also needs to perform random basis choices [1]. He also begins to count the $N_{B0}$ number of clocks before any photons are transmitted until he detects a dark count (once again he blocks the detector input), and calculates $N_{B0}$ mod 2 to determine the first basis to be used. He follows the same procedure like in the BBM92 protocol, calculating $N_{BT}$ mod 2 for each basis. Moreover, if Alice uses a heralded single photon source [21,22], she also disposes of a single photon detector that can be exploited to generate random numbers in the exact same way, both for the bit and the basis choice, just calculating $N_{AT}$ mod 4 and converting the 4-valued number into 2 random bits.

Alice and Bob can increase practical security by instead of blocking their detector inputs for just a single detection, they actually increase it to $k$ detections. In this case both of them have each a local electronic buffer of $k$ positions to further increase the number of basis choices that Eve cannot perform attacks on. One last note is that our protocol is independent of the encoding method, and can be readily applied to phase encoded systems.

Now let us look at the system from Eve's point of view. It is clear that in order for the system to be secure, she cannot predict which bases Alice and Bob will choose for each transmitted qubit. The simplest attack Eve can perform on our scheme is a Quantum Non-Demolition (QND) measurement [23] to verify in the transmission line in which time instants a single photon exists and, from the results, predict the bases that will be chosen. However, if we once again assume Eve is not inside Alice or Bob's stations, she will not be able to predict the bases, due to the random nature of the detection events. In order to guarantee security they use their local clock signals with a resolution higher than the coherence time of the single photons, thus introducing a random element Eve cannot predict, making it impossible that she can make a deterministic guess. To obtain maximum efficiency the single photons should have the same coherence time as the detector's gate window, which also lowers the local clock required operating frequency. Using such a proposal for current commercial detectors typical gate widths of 2.5 ns, the detection system for the proposal presented must be able to check if the photon arrived in the first or second halves of the gate, that is, the clock period must be smaller than 1.25 ns. The corresponding frequency of 800 MHz, although tricky to work with, is perfectly achievable in modern electronic circuits. Alice and Bob may also choose to use additional interference filters that guarantee the photons have a coherence time equal to the detection window (almost a requirement for current SPDC - Spontaneous Parametric Down-Conversion - sources that typically have bandwidths of several nm). This also offers protection against a more sophisticated attack in which Eve replaces the photons with ones with lower coherence time such that she can force them to be in the beginning or end of the detection window. In order to obtain *practical* security Alice and Bob can use a local clock with a higher resolution than the jitter of the detectors, allowing maximum efficiency while using our scheme with SPDC based sources as there is no need for filtering. This places more stringent requirements on the local clock, as it now needs to have a resolution better than tens of ps. Although unlikely to be implemented with current technology, future improvements in electronics will allow it to become a reality.

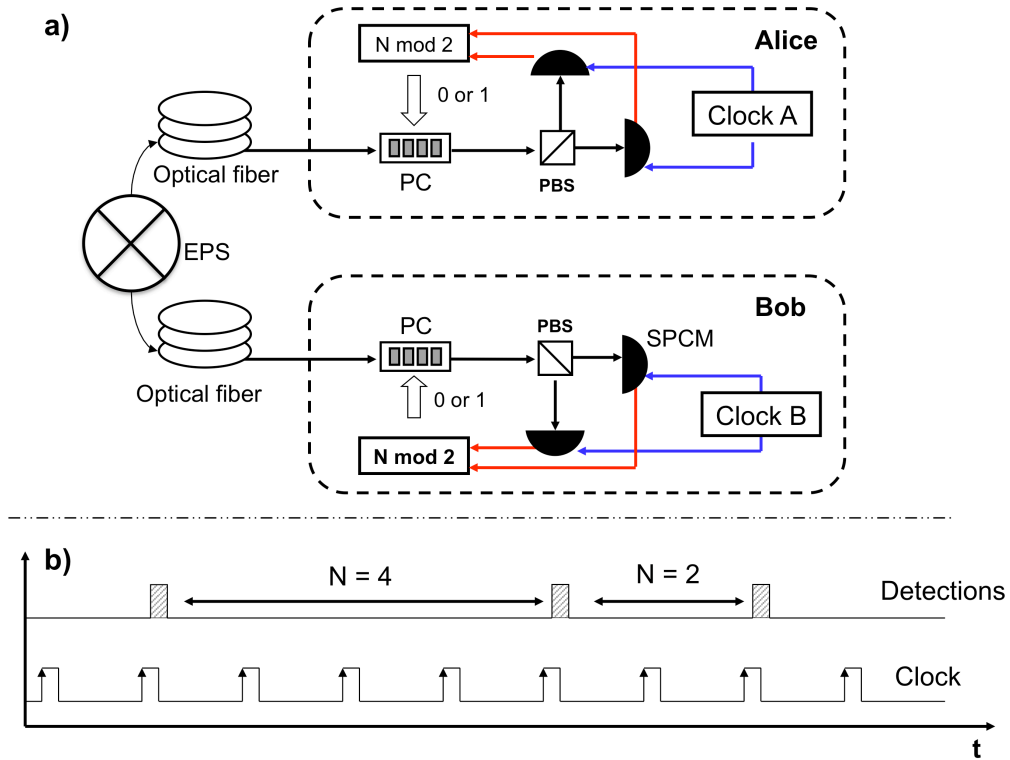

Fig. 1. a) Schematics of our proposal applied to the BBM92 protocol. Black arrows represent optical connections, while blue and red ones depict electrical cables. EPS - Entangled Photon Source; PC - Polarization Controller; PBS – Polarizing Beam Splitter; SPCM – Single Photon Counting Module. The master clock synchronizing Alice and Bob, as well as QKD electronics are omitted for the sake of clarity. b) Illustrative representation of the waveforms from the detection and clock pulses.

If we now consider the poissonian probability of existing $n$ photons inside a detection window with $\mu$ photons per pulse [24]:

$$p(n) = \frac{e^{-\mu}\mu^n}{n!} \quad (1)$$

the probability (for any given distribution) of detecting a photon on the $N_{th}$ gate window can be written as:

$$P(N) = \left[1 - p(n>0)\right]^{N-1} \cdot p(n>0) \quad (2)$$

Using Eq. (1) we can rewrite the probability of detecting a photon as:

$$p(n>0) = 1 - e^{-\mu\eta} \quad (3)$$

where $\eta$ is the detection efficiency. Eq. (3) can be interpreted simply as having a perfect non-resolving photon number detector with an external loss given by the detection efficiency. We can now use Eq. (3) to write the probability of detection as:

$$P(N) = \left(e^{-\mu\eta}\right)^{N-1} \cdot \left(1 - e^{-\mu\eta}\right) \quad (4)$$

Finally from Eq. (4) we write the probabilities of detecting photons in odd and even detection windows, summing over $N$ for both cases:

$$P_{EVEN} = \frac{1}{1 + e^{\mu\eta}} \quad (5)$$

$$P_{ODD} = \frac{1}{1 + e^{-\mu\eta}} \quad (6)$$

From these two equations we note that the probability of odd and even detection events are different. This means that the sequence generated from $N \bmod 2$ will be unbalanced in terms of zeros and ones. This bias will decrease if we employ a higher timing resolution clock, vanishing completely at the limit of infinite high resolution, thus reinforcing our choice of a local fast clock; however, even with an unbalanced sequence, post-processing procedures can be used to overcome this issue. To make sure that a higher-resolution clock will be effective at removing the bias and protect the system from Eve's attacks, the coherence time of the light source must be of the same size as the detection window as mentioned above. In addition to that, the single-photons must be emitted in a single mode, so that the photon has equal probability of being detected at the first or second halves of the gate window and is spatially indistinguishable.

The previous discussion was done assuming light with Poissonian statistics, and for many situations it is a good approximation of the statistics of the light field [1,24]. We can also extend the previous discussion to the case of light sources with thermal statistics. In this situation the probability to find $N$ photons per detection window with $\mu$ photons per window on average is [24]:

$$p(n) = \frac{\mu^\eta}{(\mu+1)^{n+1}} \quad (7)$$

Following the same line of reasoning as before, we can arrive at the following probabilities for odd and even events:

$$P_{EVEN} = \frac{1}{\mu\eta + 2} \quad (8)$$

$$P_{ODD} = \frac{\mu\eta + 1}{\mu\eta + 2} \quad (8)$$

As before the probabilities for odd and even events differ generating a biased sequence. The bias can be diminished, once again, by using a local clock with a higher resolution. Sources based on SPDC using short pump pulses emit light with thermal statistics [25], as well as SPDC sources employing Continuous-Wave (CW) pump lasers [26,27]. However, depending on the experimental conditions, these CW SPDC sources emit photons with poissonian statistics [28], as is the typical case in many experiments. Sources based on weak coherent states, like attenuated pulsed lasers, are also poissonian distributed [1]. Before performing the proof-of-principle experiment outlined below, we ran simple computer simulations to verify the validity of our idea. The simulations were performed for both thermal and poissonian light sources, with similar results showing that random numbers were generated.

## 3. Experimental demonstration

We perform a simple proof-of-principle experiment to demonstrate that our scheme can generate random numbers which would be suitable for our protocol. The single photons are generated using the process of SPDC through a $\chi^{(2)}$} non-linear crystal [13] representing the building block for an entangled single-photon source. We used a 20 mm long Periodically-Poled Lithium Niobate (PPLN) crystal pumped by a 532 nm CW Nd:YAG laser. The crystal poling is designed to provide ZZZ quasi-phasematching in order to generate correlated photon pairs at the non-degenerate wavelengths 810 and 1550 nm (signal and idler respectively), in the same polarization state as the pump, when the crystal is heated at approximately 90º C. The experimental setup is presented in Fig. 2. The pump light passes through an optical attenuator, and then goes through a half-wave plate (the SPDC process is polarization sensitive). It is then focused on the 20 mm long crystal using an achromatic doublet lens *L* with a 100 mm long focal length. It is collimated by an identical lens before the prism *P*, used to spatially split the generated beams and the pump. A bulk filter (RG 715) is placed before the fiber coupler (FC) to remove any residual pump light from the detector. An aspheric lens (f = 11 mm) is used to focus the signal beam on a standard 780 nm single mode fiber (SMF), mounted on a multi-axis translation stage. The fiber is connected to a Si Avalanche Photo Detector (Id Quantique ID100-MMF50). The output from the detector is finally connected to an A/D card (20 MSamples/s), which is attached to a personal computer to process the data. The local clock rate used in our measurement is therefore 20 MHz, corresponding to a 50 ns period, which is much longer than the detector jitter of 40 ps. In addition to that, the measured down-converted bandwidth at 810 nm with a monochromator is approximately 1 nm, yielding a coherence time of 2.18 ps. Clearly our experiment does not represent a secure QKD transmission based on the discussion from the previous section. Nevertheless it is still useful to demonstrate the validity of our idea. The optical attenuator was adjusted so that a 150 kHz detection rate was obtained with optimized coupling. The measured input optical pump power on the crystal for that rate is 9.8 mW. The experiment we perform here is a perfect representation of Alice in the case where she uses a heralded single-photon source. It also represents the basic building block of the BBM92 protocol, with the source located somewhere between Alice and Bob. Our setup can therefore be easily upgraded into a single-crystal entangled photon pair source [14,15].

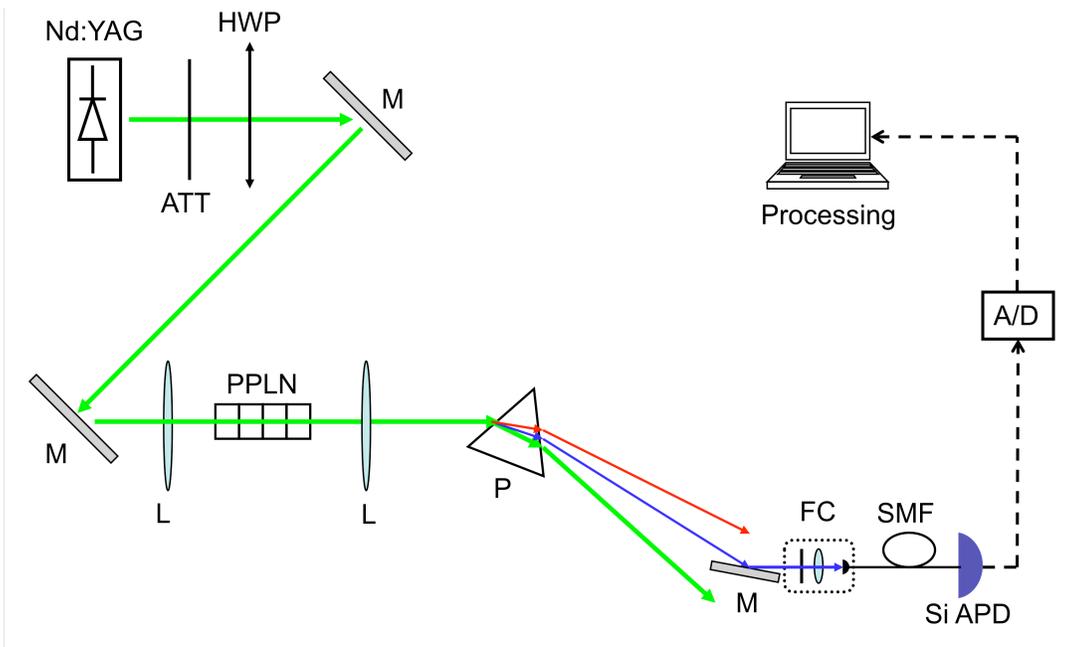

Fig. 2 Experimental setup: ATT: Optical attenuator; HWP: Half-wave plate; M: Mirror; L: Lens; PPLN: Periodically-poled lithium niobate; P: Prism; FC: Fiber coupler, here consisting of a multi-axis translation stage (not shown here), RG 715 high-pass filter, 11 mm focal length aspheric lens and fiber holder; SMF: 780 nm single mode optical fiber; APD: Avalanche photon detector; A/D: Analog to digital converter. The green, blue and red arrows represent the pump, signal and idler beams respectively. The dashed lines represent electrical cables.

A widely used standard measure of randomness is the NIST statistical test suite version 1.8 [29]. It comprises a series of 13 rigorous procedures to test random number generators [7,8,11]. We generated a sequence of 20 million bits at 150 kHz count rate, using the previously described $N$ mod 2 procedure. Our clock resolution was limited by the A/D card sampling rate (20 MHz), and it is not enough to entirely remove the bias, as discussed before. Since the NIST suite expects an unbiased sequence, we adopted a very simple procedure of flipping bit assignments at each detection in order to make it balanced. With this procedure, our bit sequence balance was increased from 99.68 % to 99.95 %. A single run of the suite requires 1 million bits, so we performed it 20 times. Each of them returns a number, called a p-value, between 0 and 1. As long as the p-value is greater than the confidence value (we used 0.01, which is the standard for random sequence tests [29]), then the sequence has passed that test. The results for the 13 tests are presented in Fig. 3, which indicates that our sequence is random with a very good degree of confidence.

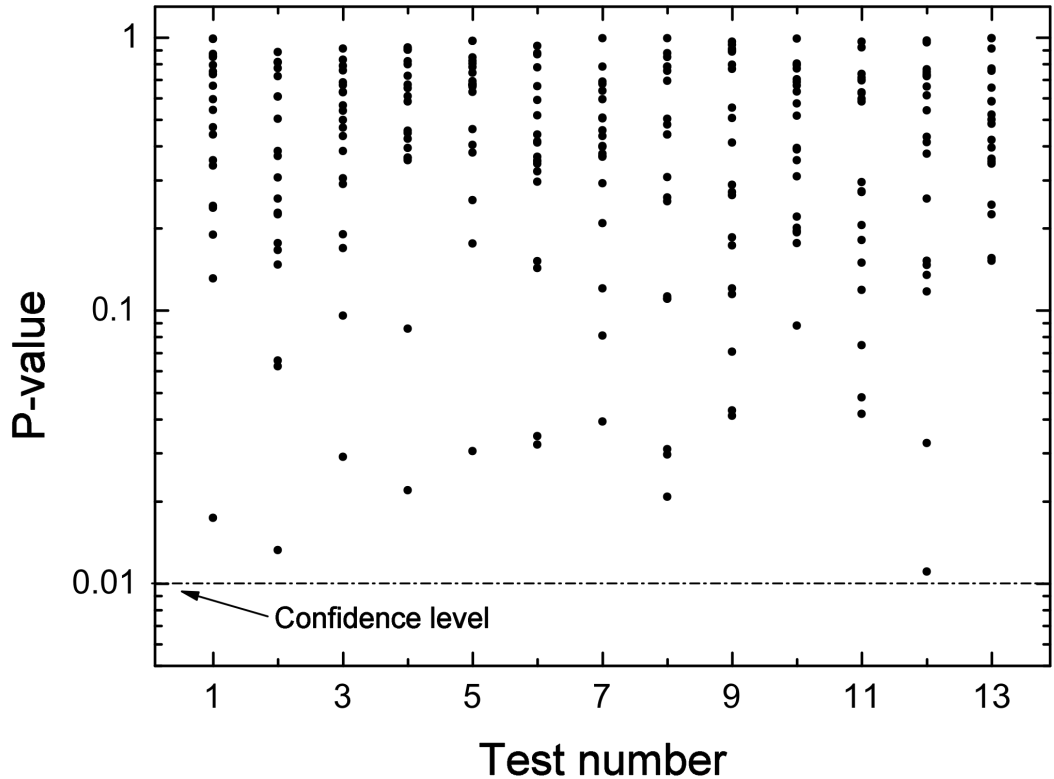

Fig. 3 P-values plotted for the NIST test suite individual tests for the 20 million bit generated sequence after bias removal. Each dot represents a run of 1 million bits for a particular test. The results are all above the confidence value for cryptography applications. The tests are: 1 – Frequency; 2 – Block frequency; 3 – Cumulative-sums forward; 4 – Cumulative-sums reverse; 5 – Runs; 6 – Longest runs; 7 – Rank; 8 – DFFT; 9 – Universal; 10 – Approximate entropy; 11 – Serial 1; 12 – Serial 2; 13 – Linear complexity.

## 4. Conclusions

We have presented a scheme to perform true random basis choices for entanglement-based QKD protocols (with a focus on BBM92), based on the hardware that is already present in any QKD system. It may also be extended to be used by Bob in the BB84 protocol, or even Alice, if she uses a heralded single photon source. Our proposal has the advantage of being readily scalable "on-the-fly" with the transmission rate without any active changes from the user, as long as high-resolution local clocks are used. Therefore, it could be used in future entangled based QKD networks [30], or any quantum cryptography system which employs a variable key rate. The protocol can be implemented with simple modifications and it replaces true RNGs for the active basis choices, decreasing the building cost of a practical QKD setup. We have shown that the generated sequence is indeed random like we expected, and supports our idea of random number generation in QKD systems. If Alice and Bob employ asynchronous clock generators with a period of at least half of the detection window, and the coherence time of the single-photons is of the same width as the window, then Eve cannot gain any information on the basis selection.

The authors wish to thank N. Lütkenhaus, N. Gisin and H. Zbinden for helpful discussions, S. Sauge and A. Karlsson for the PPLN crystal lent. Financial support is acknowledged from CNPq and FAPERJ.